\documentclass[preprint2]{emulateapj}
\shorttitle{Radio continuum observations of  VCC128} \shortauthors{Buyle, De
Rijcke, Debattista, Ferreras, Pasquali, Seth, Morelli}

\begin{document}

\title{Radio continuum observations of the candidate supermassive
black hole in the dwarf elliptical VCC128}

\author{Pieter Buyle\altaffilmark{1}, Sven De Rijcke\altaffilmark{1},
Victor P. Debattista\altaffilmark{2,3}, Ignacio Ferreras\altaffilmark{4},
Anna Pasquali\altaffilmark{5}, Anil Seth\altaffilmark{6},
Lorenzo Morelli\altaffilmark{7}}

\altaffiltext{1}{Sterrenkundig Observatorium, Universiteit Gent,
Krijgslaan 281, S9, B-9000, Ghent, Belgium; Pieter.Buyle@UGent.be,
Sven.DeRijcke@UGent.be} 
\altaffiltext{2}{Centre For Astrophysics
University of Central Lancashire
Preston, UK PR1 2HE; vpdebattista@uclan.ac.uk} 
\altaffiltext{3}{RCUK Academic Fellow} 
\altaffiltext{4}{Physics Department, King's College
London, Strand, London WC2R 2LS, UK; ferreras@star.ucl.ac.uk}
\altaffiltext{5}{Max-Planck-Institut f\"ur Astronomie, Koenigstuhl 17,
D-69117 Heidelberg, Germany; pasquali@mpia-hd.mpg.de}
\altaffiltext{6}{Harvard-Smithsonian Center for Astrophysics, 60 Garden St., MS20, Cambridge, MA 02138, US; aseth@cfa.harvard.edu}
\altaffiltext{7}{Dipartimento di astronomia, Universita' di Padova, Vic. Osservatorio 3, I-35122 Padova, Italy; lorenzo.morelli@unipd.it}

\begin{abstract}
The presence of black holes (BHs) at the centers of dwarf elliptical
galaxies (dEs) has been argued both theoretically and
observationally. Using archival HST/WFPC2 data, we found the Virgo
cluster dwarf elliptical galaxy VCC128 to harbor a binary nucleus, a
feature that is usually interpreted as the observable signature of a
stellar disk orbiting a central massive black hole. \citet{db06}
estimated its mass M$_\bullet \sim 6 \times 10^6-5\times
10^7$~M$_\odot$. One of the most robust means of verifying the
existence of a BH is radio continuum and/or X-ray emission, however
because of the deficiency of gas in dEs, radio continuum emission
forms the best option here. We have tried to detect the X-band radio
emission coming from the putative black hole in VCC128 when it
accretes gas from the surrounding ISM. While we made a positive
4$\sigma$ detection of a point source 4.63$''$ south-west of
the binary nucleus, no statistically significant evidence for emission
associated with the nuclei themselves was detected.

This implies either that VCC128 has no massive central black hole,
which makes the nature of the binary nucleus hard to explain, or, if
it has a central black hole, that the physical conditions of the ISM
(predominantly its density and temperature) and/or of the surrounding
accretion disk do not allow for efficient gas accretion onto the black
hole, making the quiescent black hole very hard to detect at radio
wavelengths.
\end{abstract}

\keywords{galaxies: evolution, galaxies: elliptical, galaxies: ISM, galaxies: fundamental parameters}

\section{Introduction}\label{intro}

The masses of the massive central black holes observed in many
galaxies exhibit a variety of scaling relations, such as the
M$_\bullet - \sigma$ relation between black-hole mass, M$_\bullet$,
and central velocity dispersion, $\sigma$ \citep{g00,fm00} and the
$v_{\rm circ} - \sigma$ relation between circular velocity, $v_{\rm
circ}$, and central velocity dispersion
\citep{f02,b03,pi05,bfg06}. The latter can be interpreted, via the
M$_\bullet - \sigma$ relation, as a relation between the mass of the
central black hole and the total mass of the host galaxy. These
correlations suggest a strong coupling between the formation of
massive central black holes and the formation of galaxies.

Literature abounds with scenarios for producing massive central black
holes. Models designed to explain the properties of QSOs in massive
galaxies typically produce central seed black holes with a minimum
mass of M$_\bullet \gtrsim 10^6$M$_\odot$, e.g. through the collapse
of a massive central gaseous disk or a supermassive stellar object
\citep{lr94,h98,sr98}. This seed can grow even more massive by feeding
from a surrounding gas disk. These models are motivated by the
necessity of producing supermassive black holes very rapidly after the
Big Bang and therefore may be biased to large seed masses.  The
gravitational collapse of a relativistic star cluster, supposedly born
during a starburst, might produce black hole seeds with masses up to
$10^4$~M$_\odot$ \citep{s04}. Scenarios for growing less massive black
holes, aptly called intermediate-mass black holes, also exist in the
literature \citep{mr01,mh02,pm02}. Their applicability seems to be
restricted to very dense systems, such as, for instance, globular
clusters or galactic nuclear star clusters, because they rely on
dynamical friction to sink stellar-mass seed black holes, possibly
originating from Pop{\sc iii} objects as suggested by \citet{mr01}, to
the center of the system on short enough timescales.

The picture becomes even more complicated if the fact that galaxies
grow via a process of hierarchical merging is taken into
account. Recently, the impact of the co-evolution of central black
holes and their host galaxies has been incorporated in semi-analytical
models (SAMs) of galaxy evolution. E.g., \citet{kh00}, \citet{m07} and
\citet{ka05} allow massive central black holes to grow by two
processes:~{\em (i)} the influx of gas into the galaxy center due to
the starburst and the strongly non-axisymmetric forces that accompany
a galaxy-galaxy merger, and {\em (ii)} merging of the black holes
after the host galaxies have merged. These SAMs can reproduce the
aforementioned scaling relations between the black-hole mass and the
properties of the host galaxy. Using a high-resolution small-volume
cosmological N-body/SPH simulation, \citet{mhbs07} show they can grow
M$_\bullet \sim 10^6$~M$_\odot$ central black holes in Milky-Way size
halos by a redshift of $z \sim 6$ by allowing central black holes to
merge if their host galaxies merge and by gas accretion during
mergers.

However, when the members of a binary black-hole system merge, there
can be a considerable beaming of the gravitational radiation emitted
during the final stages of the plunge. This results in a recoil of the
newly formed black hole. Fully relativistic calculations of black-hole
binary mergers have shown that this recoil velocity can be as high as
a few hundred~km~s$^{-1}$, depending on the holes' spin, orientation,
and orbital eccentricity \citep{p07}. This could potentially hamper
the retention of black holes in merging low-mass protogalaxies and by
extension the build-up of supermassive black holes.

Clearly, the search for massive central black holes in low-mass
galaxies is of considerable interest for the study of the symbiosis of
host galaxies and their central black holes and for constraining the
minimum mass of seed black holes. Nuclear activity in dwarf galaxies
has turned up a number of central black holes with estimated masses in
the range M$_\bullet \sim 10^4-10^6$~M$_\odot$
\citep{fh03,b04,gh04}. Dynamical modeling of ground-based stellar
kinematics of dwarf elliptical galaxies ruled out the presence of
central black holes with masses M$_\bullet \gtrsim 10^7$~M$_\odot$
\citep{g02}.

\begin{figure*}
\includegraphics[width=18cm]{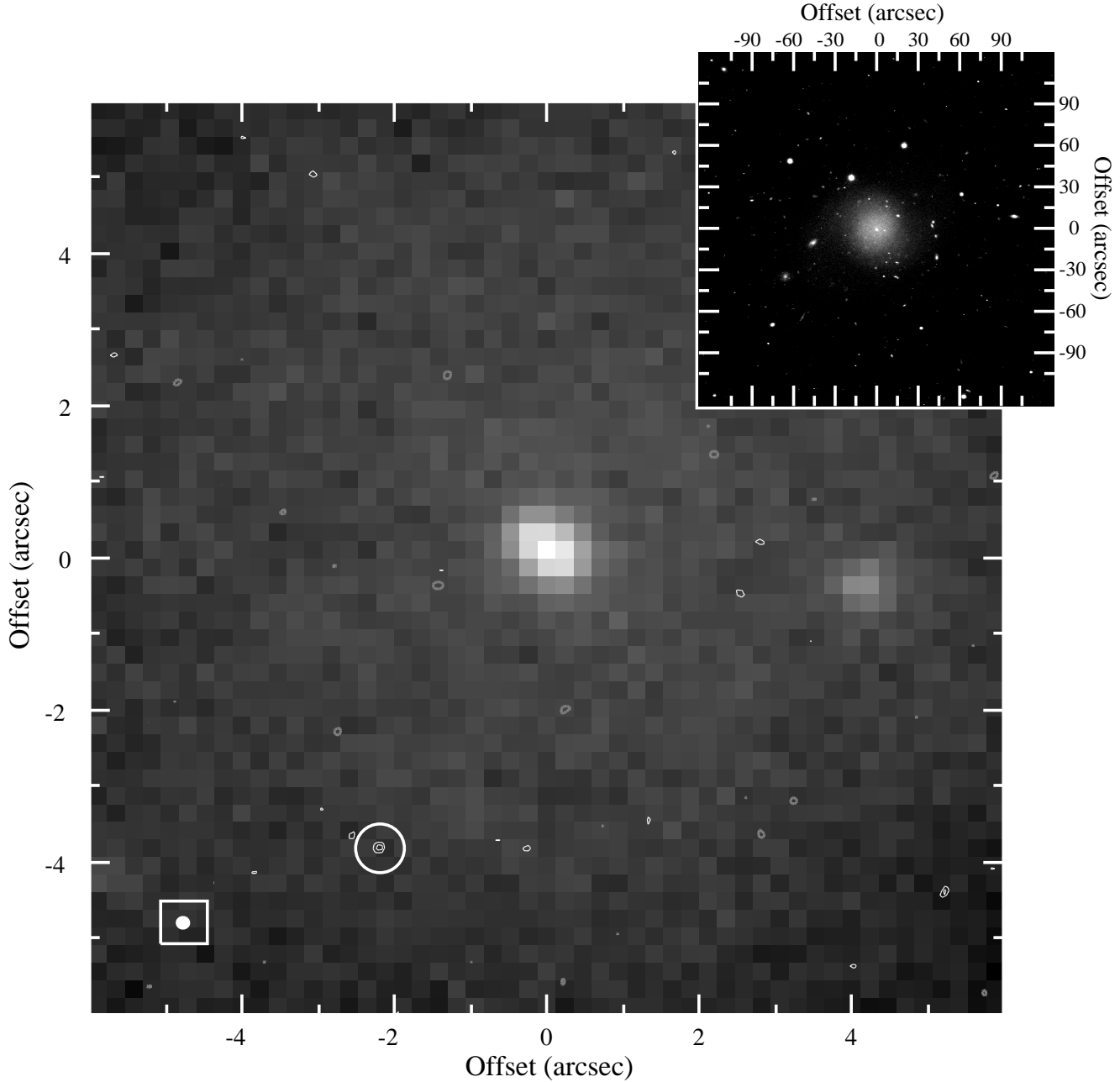}
%\caption{V-band image of VCC128. The inset shows a $12.5''\times
%  12.5''$ region centered on the binary nucleus, which is just
%  resolved in this image, overlayed with a contourplot of the VLA
%  radio observations. In increasing order, the contourlevels are
%  3$\sigma$, 4$\sigma$, 5$\sigma$, \ldots with $1\sigma$ =
%  $30\mu$Jy. The detected source, at $4\sigma$ significance, can be
%  seen SW of the center (encircled).
%\label{ima_128}}
\caption{Contourplot of the VLA radio observations on top of a V-band
  image of the center of VCC128. The figure shows the $12''\times
  12''$ region centered on the binary nucleus, which is just
  resolved in this image. The white contours correspond in increasing
  order to the contourlevels 3$\sigma$, 4$\sigma$, 5$\sigma$, \ldots
  with $1\sigma$ = $30\mu$Jy, while the gray contours correspond to
  $-3\sigma$. The detected source, at $4\sigma$ significance, can be
  seen SW of the center (encircled). The synthesized beam is shown in
  the bottomleft corner. A V-band image of entire VCC128 is shown in
  the topright corner.
\label{ima_128}}
\end{figure*}

Obtaining and modeling high-quality stellar kinematics of faint dwarf
galaxies is, unfortunately, a very time-consuming way of searching for
inactive central black holes. This prompted us to look for a more
efficient, photometry-based method. Using archival HST images, we
found the Virgo Cluster dwarf elliptical galaxy (or dE) VCC128 to
harbour a binary nucleus \citep{db06}. At the time, only two other
galaxies with binary nuclei were known: M31 and NGC4486B
\citep{l93,l96,l05}. In both cases, the binary nucleus was interpreted
as the observable signature of a stellar disk orbiting a central
massive black hole \citep{t95,l96}. This suggested that the binary
nucleus in VCC128 might also be the hallmark of a central massive
black hole.

However, the presence of a central massive black hole in VCC128 still
needs to be established by independent means. An observationally
rather undemanding method for detecting massive central black holes is
through the radio emission they produce when they accrete gas. In
Section \ref{rce}, we give a short discussion of this method, followed
by an account of our radio continuum observations of VCC128 with the
Very Large Array (VLA) in Section \ref{obs}. We end by presenting our
results and conclusions in Section \ref{results}.

\section{Radio flux density prediction for an accreting black hole} \label{rce}

In order to predict the expected radio continuum flux density of an
actively accreting BH, we need an estimate of the BH's mass. In the
case of VCC128 we can make use of the optical photometry from Hubble
of the double nuclei. By comparing the observed SED of the nuclei with
population synthesis models, we estimated the combined nuclear stellar
mass at $\sim 10^6$~M$_\odot$. Assuming that the disk to black-hole
mass ratios of M31 (M$_{\rm disk}$/M$_\bullet=0.16$) and NGC4486B
(M$_{\rm disk}$/M$_\bullet=0.019$) are typical, \citet{db06} estimated
the mass of the putative black hole in VCC128 at M$_\bullet \sim 6
\times 10^6-5\times 10^7$~M$_\odot$. As a verification they used the
Faber-Jackson relation of dEs in order to estimate the galaxy's
central velocity dispersion, $\sigma$, at $\sigma \sim
35-65$~km~s$^{-1}$. This result is in good agreement with the
empirical scaling M$_\bullet - \sigma$ relation \citep{mf01,w06}.

\citet{mhd03} observationally derived a fundamental plane relation
between 5~GHz radio luminosities $L_{\rm cont}$, X-ray luminosities
$L_X$ and BH masses M$_\bullet$ of Galactic and extragalactic
BHs. Unfortunately, an X-ray luminosity does not exist for VCC128.
Still, the fundamental plane relation and its projections provide us
with a useful tool to estimate the radio continuum flux density
produced by VCC128 if an actively accreting BH were present (see
\citet{bash08} for a similar approach). From Fig. 2 in \citet{mhd03},
for M$_\bullet \gtrsim 6 \times 10^6$~M$_\odot$ one would expect a
5~GHz radio luminosity well above $L_{\rm cont} \sim 10^{35}$
erg~s$^{-1}$.

An alternative method to obtain the radio flux $F_{\rm cont}$ without
knowing $L_X$ is proposed by \cite{ma04}, assuming a flat radio
spectrum and the Fundamental-Plane relationship between radio
luminosity, X-ray luminosity, and black hole mass
\citep{mhd03,fa04,dbd06}. For a given gas accretion rate onto the
black hole, expressed as a fraction of the Bondi accretion rate which
depends on the gas density and temperature, the X-ray luminosity can
be calculated, assuming that 10~\% of the rest mass energy of the
infalling matter is converted into radiation. 
%, and that the total
%luminosity $L_{\rm tot}$ is related to the X-ray luminosity as given
%by $L_{\rm tot} = 6\times 10^{-3} L_X^{0.5} + L_X$ (using the
%Eddington luminosity as unit). This relation takes into account the
%energy of a relativistic jet \citep{fe03}. This way, the radio
%continuum flux $F_{\rm cont}$ can be expressed as a function of the
%black hole mass $M_\bullet$, the distance $d$, and the gas density $n$
%and temperature $T$.
Unfortunately, no H$\alpha$ imaging or 21~cm radio line emission data
are available for VCC128, leaving us without direct constraints on the
gas density and temperature or even an indication that gas is at all
present in this dwarf galaxy.

Thus, it is in principle possible to detect massive central black
holes and to estimate their masses (or place broad upper limits
thereon), allowing for a large uncertainty due to the combination of
several empirical scaling relations and the unknown properties of the
accreting gas, using straightforward and relatively short radio
observations. We wish to stress that the detection of a radio
continuum signal from the center of VCC128 would provide strong
evidence for the presence of a central black hole, independent of the
reliability and accuracy of the methods with which we can estimate its
mass.

\section{Observations and data reduction}\label{obs}

We observed VCC128 with the Very Large Array (VLA) in New Mexico (USA)
around midnight of June 28, 2007 (project number AB1248). The VLA is
being replaced gradually by the Expanded VLA (EVLA) and, therefore, at
the time of our observations consisted of 16 antennas. The radio
continuum emission in a region around the center of the galaxy was
mapped in the X-band (8.6~GHz), which is at the VLA's maximum
sensitivity. The VLA was in its A configuration at that time, yielding
a spatial resolution of $0.24''$ corresponding to 19~pc at the
distance of VCC128 (16.5~Mpc). The total project time was $3~h$. At
startup, the flux calibrator 1331+305 was observed for 7 minutes,
followed by alternating observations of the phase calibrator 1222+042
(with an angular separation of 5.5$^\circ$ from VCC128) with a 2
minute exposure time and the galaxy VCC128 with a 15 minute exposure
time. This gives a total on-source integration time of 129
minutes. Standard flagging and calibration of the $u-v$ data was
performed with the Astronomical Image Processing Software (AIPS). We
preferred to use MIRIAD \citep{sault95} to derive a natural weighted
map. This map was cleaned (1000 iterations) and yields a final image
beam size of 0.19$'' \times $0.18$''$.

\section{Results and conclusions}\label{results}

We created a radio continuum map of the central $12''\times 12''$
region of VCC128 with an rms noise of $30~\mu$Jy/beam. This map is
presented in Fig. \ref{ima_128}, overplotted on a V-band image of
VCC128 obtained with FORS1 mounted on the VLT (ESO program
079.B-0632(B)). The data reduction and analysis of the optical imaging
of VCC128 will be reported in detail elsewhere. A 4$\sigma$ detection
of a point source was made at R.A. $12h~14m~59.66s$, DEC. $+9^\circ~
33'~50.94''$ (J2000). This is 4.63$''$ SW of the binary
nucleus. The source has a flux density of $1.66 \pm
0.05$~mJy. Unfortunately, no emission was found at the optical
position of the nucleus.

Since we do not detect any radio emission coming from the center of
VCC128, we can only place a $S_\nu \approx 90~\mu$Jy $3\sigma$ upper
limit on the radio continuum emission of the putative central black
hole. Assuming a flat radio spectrum, this corresponds to a $3\sigma$
upper limit on the 5~GHz radio luminosity of $L_{\rm cont} = \nu L_\nu
\approx 1.5 \times10^{35}$~erg~s$^{-1}$. This is lower than the expected
radio luminosity of an actively accreting BH with a mass M$_\bullet
\gtrsim 10^6$~M$_\odot$.

We conclude that either {\em (i)} VCC128 does not contain a central
massive black hole or, alternatively, {\em (ii)} VCC128 might contain
a central massive black hole with a mass of the order of a few
$10^6$~M$_\odot$ but the gas near the galaxy center is either too hot
or too rarefied or conditions near the black hole are unsuitable for
efficient accretion. In the former case, one needs to rethink the
nature of the binary nucleus. As detailed in \citet{db06}, all other
explanations for the binary nucleus other than it being the
observational signature of a stellar disk rely on rather contrived
chance configurations. 

%Given the very shallow gravitational potential
%of VCC128 it would then be very unclear how this disk can remain
%stable at the galaxy center without some central, unseen mass
%concentration.

%Although no detection is made, we can still use the M$_\bullet-{\rm
%  F}_{\rm cont}$ relation of \citet{ma04} to derive very broad
%constraints on the physical conditions near the putative BH in VCC128
%that would prevent it from emitting more radio continuum emission:
%\begin{itemize}
%\item the gas density must be below $\sim 0.005$~cm$^{-3}$, for
%$T=10^4$~K and an accretion rate of $10^{-3}$ times the Bondi rate,
%\item the accretion rate must to well below $10^{-5}$ times the Bondi
%rate, for $T=10^4$~K and $n=0.1$~cm$^{-3}$,
%\item the temperature must be higher than $\sim 10^5$~K, for
%$n=0.1$~cm$^{-3}$ and an accretion rate of $10^{-3}$ times the Bondi
%rate.
%\end{itemize}
%In short, our VLA radio continuum observations of VCC128 did not yield
%a detection of an accreting massive central black hole with a mass of
%%the order M$_\bullet \sim 6 \times 10^6-5\times 10^7$~M$_\odot$ (as
%claimed by \citet{db06}) in this Virgo cluster dwarf elliptical
%galaxy. Our observations however are not conclusive as many different
%physical conditions are involved in the determination of the accretion
%rate of a BH.

\acknowledgements PB and SDR are post-doctoral fellows of the Fund for
Scientific Research - Flanders, Belgium (F.W.O.). LM is supported by
grant (CPDR061795/06) by Padova University. The Very Large Array is a
facility instrument of the National Radio Astronomy Observatory, which
is operated by Associated Universities, Inc. under contract from the
National Science Foundation.

\end{document}